\documentclass[12pt]{article}
\usepackage{epsfig}
\usepackage{mathtools}
\usepackage{tabulary}
\usepackage{array}
\usepackage[version=4]{mhchem}
\usepackage{float}
\usepackage[utf8x]{inputenc}
\usepackage{amsmath}
\usepackage{caption}
\usepackage{multirow}
\usepackage{geometry}
\usepackage{pdflscape}
\usepackage{booktabs}

\def\beq{\begin{equation}}
\def\eeq#1{\label{#1}\end{equation}}
\def\eeqn{\end{equation}}

\def\beqa{\begin{eqnarray}}
\def\eeqa#1{\label{#1}\end{eqnarray}}
\def\eeqan{\end{eqnarray}}

\let\bar=\overbar

\def\Dslash{\not{\hbox{\kern-4pt $D$}}}
\def\dslash{\not{\hbox{\kern-2pt $\del$}}}

\def\msb{{\bar{\ssstyle M \kern -1pt S}}}

\def\Title#1{\begin{center} {\Large {\bf #1} } \end{center}}

\begin{document}

\Title{Studies of Bi-layers Growth Mechanism of Silver Bromide Molecular Clusters Prepared Via Electroporation of Vesicles and Quantum Confinement Effects Applications of Molecular Clusters}

\bigskip\bigskip


\begin{raggedright}  

{\it John Hongguang Zhang\index{Zhang, J.}\\
Altamont Research\\
Altamont, NY12009}
\bigskip\bigskip
\end{raggedright}

{\bf Abstract:}  Quantum confine system in which interactions can be tuned over a vast range
may enable profound changes in the way we understand and explore physics of the microscopic
realm. For example, it may lead to previously unknown phases of matter and aid in the discovery of new phenomena. Our previous work show that in the molecular cluster regime, the band blue shift associated with cluster growth can be understood by a model that assume electrons are confined to a spherical potential well and the clusters are made of some basic units. A formula is given for the lowest excited electronic state energy. This expression contains an electron-hole-pair (EHP)  delocalization constant $ \zeta $ as an adjustable parameter which, however, can be anchored to a definite value through the known transition energy at the spectra turn-around point. We also proposed  symmetry and probability principles in molecular cluster growth range to explain the molecular cluster electron absorption spectra turn-around phenomena and unusual isotopic properties of small silver bromide clusters. In this paper, based on systematical review of Quasi-Elastic Light Scattering (QELS), Fourier-transform infrared spectroscopy (FTIR) and Direct Laser Desorption Mass Spectra (DLD-MS) experiments of silver bromide clusters prepared via the electroporation of vesicles, we show how the symmetry and probability principles in molecular cluster growth range can be used to explain the lager silver bromide molecular cluster formation in a bilayer formation mechanism. The $\lambda$ curve of $\Delta E$ vs sphere size $R$ had been proved to be a right curve to describe the molecular cluster quantum confinement behavior. We also defined Electron Delocalized Status (EDS), $ \zeta=1 $ ,  Electron Localized Status (ELS), $\zeta=0$  and EDS/ELS switch or quantum confinement switch (QCS) in the quantum confine system. These studies pave the theoretical and technical ways for advanced device technology continue shrink and new concept device generation in the atomic and molecular cluster size range.

\section{Introduction}

One of the most important property of molecular clusters is the massive changes in physical properties as a function of the size ~\cite{Rossetti}. This is usually called Quantum Size Effects (QSE)~\cite{Onushchenko}. Nanomaterials exhibit physical and chemical properties very different from those of their bulk counterparts, often resulting from enhanced surface interactions or quantum confinement~\cite{Scholl} . In the past decades, QSE had found wide range applications including superconductivity ~\cite{Guo}, visible light emission~\cite{Cullis} , quantum dot photodetectors~\cite{Konstantatos} , and industrial catalysts~\cite{Behrens} .  QSE study is also very significant to fundamental science since the energy structure in the molecular and atomic stage is well defined by quantum mechanics and in the bulk crystal stage by band theory. However in the intermediate stage, the energy structure has not yet been well modeled~\cite{Zhang2019A,Zhang2019B}. There are many open questions in this stage and new phenomena are sure to be discovered in the future ~\cite{Alivisatos,Alivisatos1,Chen1997A} .

  The growth of silver halide particles has been the subject of many studies in conjunction with their application in photographic processes~\cite{James} and recently molecular devices~\cite{Zhang2017A, Zhang2019C}. The growth mechanism is particularly important in such a process, since the sensitivity of the photographic material depends on its grain size as well as its crystal structure~\cite{James,Tadaaki,Ehrlich}. The QSE properties of AgBr nanoparticles have been studied intensively in several ways so far. They are size dependence of exciton energies~\cite{Chen1994A,Masumoto}, size dependence of exciton lifetime~\cite{Johannson,Marchetti,Kanzaki, Freedhoff}, size dependence of Raman scattering cross section~\cite{von,Scholle2,Vogelsang}     and the size dependence of the UV absorption spectra ~\cite{Tanaka,He,Zhang1997A}.  Although the QSE of nano AgBr semiconductor quantum dots with diameters exceeding 3 nm have been well characterized ~\cite{Brus,Chen1994B}, QSEs of AgBr nanoparticles below this size are poor understood. Experimentally synthesize the clusters in this smaller size regime has been extremely challenging since the spontaneous self-aggregation of the constituent molecules must be controlled and halted at the quantum size level~\cite{Zhang2000A}  .  Furthermore, optical detection is hampered by these particles’ diminishing scattering and absorption intensities. At this small particle size range, absorption measurements can give information without serious disturbance caused by light scattering, but the challenge is to detect the weak absorption of these small particles~\cite{Tanaka1} . For example, pulse radiolysis was used to study the formation of the AgBr monomer and the $(AgBr)_2$ dimer in solution. $AgBr$  monomer absorption at 295 nm was successfully observed shift to blue to form dimer $(AgBr)_2$  at 285 nm ~\cite{Zhang1997A}. However, the absorption intensity decreasing and   the absorption peak broaden made it difficult to observe the trimer and tetramer with this method.  Method of preparation of AgBr quantum dots via electroporation of vesicles let us successfully synthesized the $AgBr$  clusters from 5Å to 2nm.  For the first time, we observed the entire blue-shift (274nm to 269nm) followed by red-shift (269nm to 273nm) of the absorption band that is associated with the growth of the silver bromide clusters. The turn-around point is at 269 nm ~\cite{Zhang2000A}. The molecular and electronic structures and UV absorption spectra of neutral $(AgBr)_n$ clusters (n =1-9) were investigated in both the gas phase and in a dielectric medium by Ab initio density functional theory (DFT). The computational results parallel the experimental trends and predict that the maximum blue shift occurs at the trimer or the tetramer~\cite{Zhang2000B} . By using direct laser desorption mass spectra, we successfully observed $(Ag_2Br)^+$ , $(Ag_3Br_2)^+$ , and $(Ag_4Br_3)^+$ clusters in the same three samples that were used in the UV absorption experiments. For the first time, we experimentally confirm that the turn-around point cluster structure is $(Ag_3Br_2)^+$. This is very close to our theoretical prediction ~\cite{Zhang2000C,Zhang2012A}. In our previous work~\cite{Zhang2019B},we use symmetry and probability principles in molecular cluster growth range to explain why the cluster UV absorption have such wide band and it takes so long time to observe the band shift, why DLD-TOF-MS spectra show an unsymmetrical cluster finger peaks for small clusters and why the large clusters have much lower DLD-TOF-MS spectra intensity. In this paper, we systematical reviewed  Quasi-Elastic Light Scattering (QELS), Fourier-transform infrared spectroscopy (FTIR) and Direct Laser Desorption Mass Spectra (DLD-MS) experimental results of silver bromide clusters prepared via the electroporation of vesicles, and again use the symmetry and probability principles in molecular cluster growth range to explain the lager silver bromide molecular cluster formation in a bilayer formation mechanism. The experimental discussion and summaries can be used to  make the cation, anion and neutral molecular clusters via the electroporation of vesicles and this paves the way to form the molecular cluster source for the selective molecular cluster deposition technology~\cite{Zhang2018A,Zhang2015A}. The $\lambda$ curve of $\Delta E$ vs sphere size $R$ had been proved to be a right curve to describe the molecular cluster quantum confinement behavior. We also defined Electron Delocalized Status (EDS),  Electron Localized Status (ELS), and EDS/ELS switch or quantum confinement switch (QCS) in the quantum confine system. These studies pave the way for advanced device technology continue shrink and new concept device generation in the atomic and molecular cluster size range.

\section{Studies of Bi-layers Growth Mechanism of Silver Bromide Clusters Prepared Via Electroporation of Vesicles }

\subsection{Quasi-Elastic Light Scattering (QELS) Studies}
The mean hydrodynamic diameter of the vesicles was determined at $25^\circ$C by QELS using a Brookhaven BI-200SM multiangle goniometer in conjunction with an argon ion laser light source (514.5 nm, 20-80 mW) and a 72-channel BI-2030 digital correlator.  The performance of the instrument was evaluated by calibration using a polystyrene latex standard of 200 nm diameter.  All measurements were carried out at a 90 degree scattering angle.  The QELS data were analyzed by the nonnegative least-squares method.

\begin{figure}[H]
\begin{center}
\epsfig{file=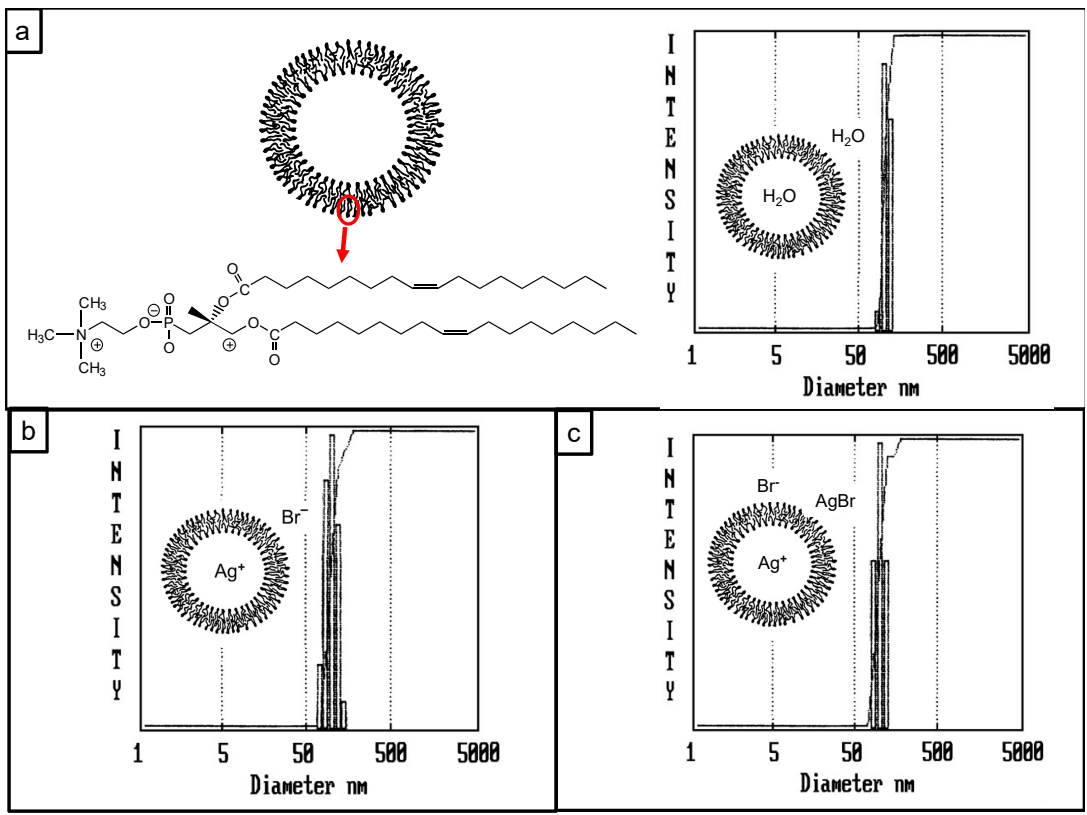,height=4.8in}
\caption{ (a) Left top is a schematic representation of the cross section of a unilamellar bilayer vesicle with silver ion inside and bromide ion outside. Left bottom is the chemical structure of DOPC. Right is the size distribution of pure DOPC vesicles in water. (b)The size distribution of DOPC vesicles containing entrapped $Ag^+$ ions, and $Br^-$ ions in the bulk prior to electroporation. (c)The size distribution of DOPC vesicles containing entrapped $Ag^+$ ions, and $Br^-$ ions in the bulk after electroporation.}
\label{fig:figure1}
\end{center}
\end{figure}

The phospholipid surfactant dioleoylphospahatidylcholine (DOPC, from Avanti Polar Lipids) was used without further purification.  After evaporating the chloroform solvent and drying under reduced pressure, large multilamellar vesicles (MLV) were prepared by hydrating the dry lipid film with a 0.01 M aqueous solution of the molecule to be encapsulated ($AgClO_4$, Fluka, 99.9\%) through mixing (Vortex-2 Genie) for about 10 min at room temperature.  The resulting solution of MLV with $Ag^+$ both entrapped and in the bulk medium had a lipid concentration of 4 mg/mL.  To prepare unilamellar vesicles, the MLV suspension was extruded five times (Extruder, Lipex Biomembranes) through two stacked polycarbonate filters of pore size 200 nm under nitrogen pressure of up to 3.4 atm. 

Figure~\ref{fig:figure1}(a) left show the phospholipid surfactant DOPC vesicle diagram and chemical formula. Figure~\ref{fig:figure1}(a) right show that at $25^\circ$C the size distribution of pure DOPC  vesicles in water.  The size distribution of DOPC vesicles containing entrapped $Ag^+$ ions, and $Br^-$ ions in the bulk prior to electroporation are shown in Figure~\ref{fig:figure1}(b) and after electroporation are shown in Figure~\ref{fig:figure1}(c). The mean hydrodynamic diameters of them are 132nm, 136nm and 139nm respectively.  We found that both the $Ag^+$ metal ions and the $(AgBr)_n$ cluster affect little on the size of DOPC vesicles. These results are similar to what we obtained before~\cite{Zhang2000B} and are common since PC vesicles are usually stable against aggregation in the presence of multivalent metal ions because of their electroneutrality~\cite{Minami}.

\subsection{FTIR Studies}   
To study the interactions of PC vesicles with metal ions or AgBr clusters by IR. It is very important to get the dry films.   To get the dry films for IR experiment, we used double polished silicon chips. These chips have area $2cm \times 1cm $ and are transparency to IR light.  They were emerged in the sample solution, and then evaporate the water solvent in a Rotavac under aspiration. After evaporating the solvent and drying under reduced pressure, the hydrated films were formed on the surface of double polished silicon chips. The silicon chips were then moved to the dry filter papers in a desiccator carefully without touch the polish silicon surface which had the sample film to be detected by the IR. The desiccator is connected to a vacuum pump. We pump the desiccator 30 minutes and then let sample dry in the vacuum desiccator about 4 hours before doing the FTIR experiments. The FTIR spectra were recorded by a Bruker VECTOR22 spectrameter, equipped with a DTGS detector in the double sided, forward-backward acquisition mode. Each spectrum was collected at a nominal $4 cm^{-1}$ resolution.

The IR studies contain three parts. The first part describes the possible cation ions interaction with the unilamellar DOPC vesicles. The second part describes the possible anion ions interaction with the unilamellar DOPC vesicles. The third part describe the IR spectra of unilamellar DOPC vesicles containing entrapped $Ag^+$ ions, and $Br^-$ ions in the bulk prior to and after electroporation.

\newgeometry{margin=2.0cm}

\begin{landscape}
{ 
\renewcommand{\arraystretch}{2.6}

\begin{table}[p]
\begin{center}

\begin{tabular}{p{3.8cm}p{2.6cm}p{2.6cm}p{2.6cm}p{2.6cm}p{3.6cm}p{2.6cm}} \hline 
& \multicolumn{6}{c}{$Peak Frequency(cm^{-1})$} \\ \cline{2-7} 
Model Assignment  & $DOPC$ & $DOPC/Ag^+$ & $DOPC/F^-$ & $DOPC/K^+$&$ DOPC/Ag^+Br^-$& $DOPC/AgBr$\\ \hline

$\nu_{as}(CH_2)$ &  2925s  & 2925s  & 2925s & 2925s & 2927s& 2925s\\ \hline
$\nu_{s}(CH_2)$ &  2855s  & 2854s  & 2854s & 2854s & 2854s & 2854s\\ \hline
$\nu_{s}(C=O)$ &  1738s  & 1739s  & 1738s & 1736s & 1736s & 1736s\\ \hline
$\delta(CH_2)_n$ &  1465m  & 1467m  & 1467m & 1466m & 1418m & 1466m\\ \hline
$\nu_{as}(PO_2 ^-)$ &  1244s  & 1210m  & 1245s & 1242s & 1236s & 1239s\\ \hline
$\nu_{s}(C-O-C)$ &  $-$  & $-$  & 1174w & 1174w & 1174w & 1174w\\ \hline
$\nu_{s}(PO_2 ^-)$ &  1090s  & 1084s  & 1091s & 1090s & 1090s & 1090s\\ \hline
$\nu_{s}(CO-O-C ^+)$ &  1066m  & 1065m  & 1065m & 1065m & 1065s & 1065m\\ \hline
$\nu_{as}(C-N^+-C)$ &  970m  & 972m  & 970m & 969m & 971m & 971m\\ \hline

\end{tabular}
\caption{Assignments of Selected Infrared Modes of DOPC vesicles,$DOPC$,  $DOPC/Ag^+$ , $DOPC/F^-$ ,  $DOPC/K^+$, $ DOPC/Ag^+Br^-$ and $DOPC/AgBr$. s-intensity strong, m-intensity middle, and w-intensity weak. }
\label{tab:IR}
\end{center}
\end{table}

} 
\end{landscape}
\restoregeometry

\begin{figure}[H]
\begin{center}
\epsfig{file=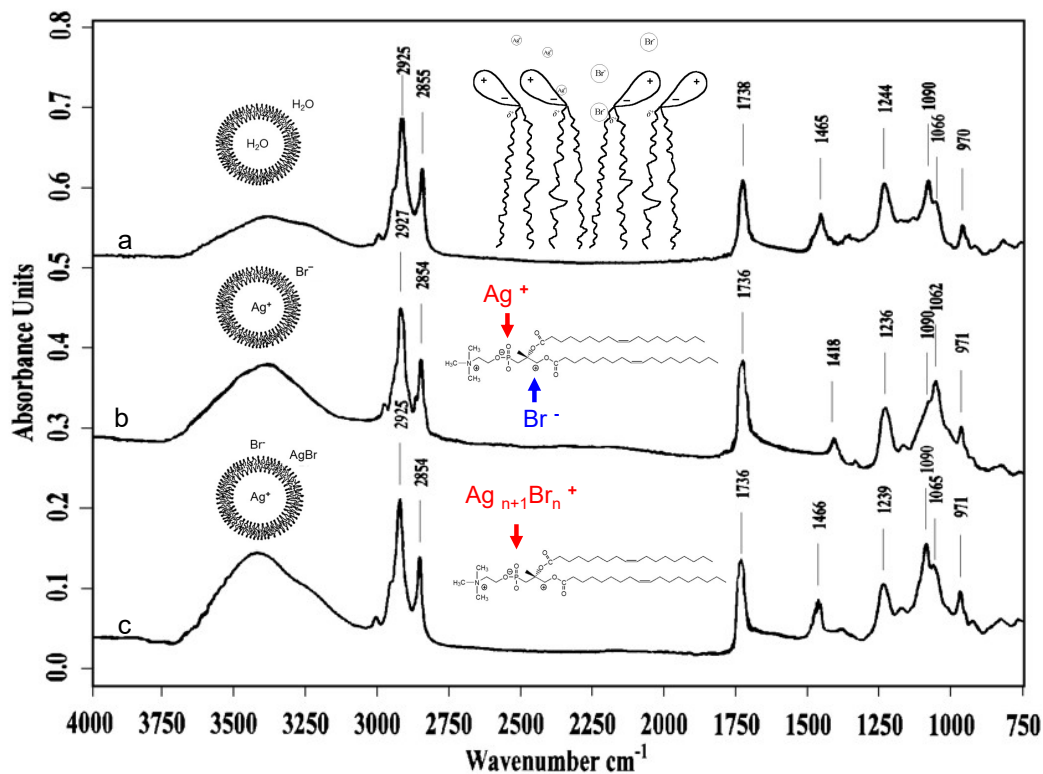,height=4.8in}

\caption{ IR spectra of (a) pure DOPC vesicles in water, (b) DOPC vesicles containing entrapped $Ag^+$ ions, and $Br^-$ ions in the bulk prior to electroporation, (c)DOPC vesicles containing entrapped $Ag^+$ ions, and $Br^-$ ions in the bulk after electroporation.}
\label{fig:figure2}
\end{center}
\end{figure}

A summary of assignments of selected infrared modes of DOPC vesicles,$DOPC$,  $DOPC/Ag^+$ , $DOPC/F^-$ ,  $DOPC/K^+$, $ DOPC/Ag^+Br^-$ and $DOPC/AgBr$ had been shown in Table~\ref{tab:IR}. To study these interactions, the IR of pure unilamellar DOPC vesicles were studied first. The pure unilamellar DOPC vesicle solution has a lipid concentration 2 $mg/ml$. The spectra are very similar to that for the solid state DOPC~\cite{Korchowiec}. The spectra has been shown in Figure~\ref{fig:figure2}(a).  The assignment of the vibrational peaks  had been listed in Table~\ref{tab:IR} according to references~\cite{Chung,Meuse}. Among these peaks, the most interesting ones can be divided into two parts. The first part are the vibrational bands related to the DOPC choline group~\cite{Akutsu1981,Akutsu1986}.  These are negative center $PO_2^-$ vibrational models $\nu_{as}(PO_2 ^-)$ at 1244 $cm^{-1}$ and $\nu_{s}(PO_2 ^-)$ at 1090 $cm^{-1}$ . Positive center $C-N^+-C$ vibrational model $\nu_{as}(C-N^+-C)$  at 970  $cm^{-1}$ and partial positive center $CO-O-C ^+$  vibrational model $\nu_{s}(CO-O-C ^+)$ at 1066  $cm^{-1}$. It is reasonable that when the cation ions attack the choline group, the vibrational bands that related to the negative center $PO_2^-$  vibrational models $\nu_{as}(PO_2 ^-)$ at 1244  $cm^{-1}$ and $\nu_{s}(PO_2 ^-)$  at 1090  $cm^{-1}$ will change. When the anion ions attack the choline group, the vibrational bands that related to the positive center $ C-N^+-C$ vibrational model $\nu_{as}(C-N^+-C)$   at 970  $cm^{-1}$ or partial positive center $CO-O-C ^+$ vibrational model  $\nu_{s}(CO-O-C ^+)$ at 1066  $cm^{-1}$ will change. These are exactly what we see in the following text. The second part is the bending vibrational mode related to the two $(CH_2)_n$ tail of DOPC $\delta(CH_2)_n$ at 1465  $cm^{-1}$. This band should very sensitive to those particle attachments on the tails. This will cause the bending vibration slow down and even vanished.

To characterize the interaction of cations with the surface of DOPC vesicles, the 0.01M $AgClO_4$ aqueous solution was used to hydrate the dry DOPC lipid film to get a lipid concentration 2 mg/ml. The resulting large multilamellar vesicles (MLV) solution were then changed to unilamellar vesicle solution according to the procedure described in the experimental section. This sample solution were then used to prepare IR samples. The IR spectra of this sample has been shown in Table~\ref{tab:IR}. From Table~\ref{tab:IR}, we can see that the vibrational bands that correspond to the negative center $PO_2^-$  vibrational models $\nu_{as}(PO_2 ^-)$  at 1244  $cm^{-1}$  and $\nu_{s}(PO_2 ^-)$ at 1090  $cm^{-1}$  had moved to 1212  $cm^{-1}$  and 1084  $cm^{-1}$  respectively. Also the intensity of asymmetric streching vibration decreased and the symmetric streching vibration increased compare to the IR spectra of pure DOPC vesicles. These wave number movement and intensity changes can be understood by the attraction of positive silver ions on the negative center $PO_2^-$ . Another interesting thing is that the positive center $C-N^+-C$ vibrational model $\nu_{as}(C-N^+-C)$ at 970  $cm^{-1}$  , partial positive center $CO-O-C ^+$ vibrational model $\nu_{s}(PO_2 ^-)$  at 1066  $cm^{-1}$  and the bending vibrational mode related to the two $(CH_2)_n$ tail of DOPC $\delta(CH_2)_n$ at 1465  $cm^{-1}$  are kept unchanged within the resolution range. They are 971  $cm^{-1}$ , 1065  $cm^{-1}$ , and 1467  $cm^{-1}$  respectively. These mean that the positive center $C-N^+-C$, partial positive center $CO-O-C ^+$ may not been attacked by the anion ions and the two $(CH_2)_n$ tail may not been attached with particles. Another characterization of the positive ions interaction with the vesicle surface is that the intensity of the negative center  $PO_2^-$  vibrational models $\nu_{s}(PO_2 ^-)$ at 1084  $cm^{-1}$ larger than that of the partial positive center$CO-O-C ^+$ vibrational model $\nu_{s}(CO-O-C ^+)$ at 1065  $cm^{-1}$ . Compare to the obvious interaction between $Ag^+$ ions and DOPC vesicles, it was found that $K^+$ ions has little interaction with the DOPC vesicles. This was done by substituting 0.01M $AgClO_4$ aqueous solution with 0.01M $KBr$ aqueous solution to hydrate the DOPC dry film to get the 2 mg/ml lipid solution. The IR spectra peaks intensity and position had been listed in Table ~\ref{tab:IR}. We can see that the $K^+$ ions have little effects on the DOPC spectra. One possible reason for this is that $K^+$  ions have the ionic radius 151 pm which are much larger than that of $Ag^+$ ions, 115 pm. Thus $Ag^+$ ions will be more easier to attack the surface of DOPC and have much larger interactions with DOPC than $K^+$ ions. Similar spectra characterizations were also found in the IR studies of the behavior of the single spreading mono layers of DOPC on sub phases containing $Na^+$ or $Ca^{2+}$ ions~\cite{Korchowiec}.  If the ionic radius is an important effect for the ions-vesicle interaction, one might expect to see the anion $–$ DOPC vesicle interaction by chose compound which has small anionic radius and large cationic radius. $CsF$ compound was chosen for this purpose, which $Cs^+$ has the ionic radius 174 pm and $F^-$ has the ionic radius 133 pm. 
To characterize the interaction of $F^-$ ions with the surface of unilamellar DOPC vesicles, the 0.01M $CsF$ aqueous solution was used to hydrate the dry DOPC lipid film to get a lipid concentration 2 mg/ml. The resulting large multilamellar vesicles (MLV) solution were then changed to unilamellar vesicle solution according to the procedure described in the experimental section. This sample solution were then used to prepare IR samples. The IR spectra peaks intensity and position had been listed in Table ~\ref{tab:IR}. Unfortunately, compare to pure unilamellar DOPC IR spectra,  there are little change which means that little interaction between $F^-$ anions and DOPC vesicles. The possible reason will be discussed.

Now let us concentrated on the IR spectra of DOPC vesicle solution of concentration 2 mg/ml with 0.01M $Ag^+$ ions inside and 0.01M $Br^-$ outside vesicle prior to Figure~\ref{fig:figure2}(b) and after Figure~\ref{fig:figure2}(c) electroporation. These will let us get some insight about the AgBr cluster$-$DOPC interaction. The procedure of making this kind of solution had been described in the experimental section. From Figure~\ref{fig:figure2}(b), we can see that the vibrational bands that correspond to the negative center $PO_2^-$   vibrational models $\nu_{as}(PO_2 ^-)$  at 1244  $cm^{-1}$  had moved only to 1236 $cm^{-1}$  this time and $\nu_{s}(PO_2 ^-)$ at 1090 $cm^{-1}$ kept unchanged. Also the difference between the intensity of asymmetric stretching vibration and the symmetric stretching vibration just increased a little compare to the IR spectra of pure DOPC vesicles. These slight change in wave number and intensity could be understood by the attacking of positive silver ions inside the vesicle on the negative center $PO_2^-$ . The positive center $C-N^+-C$ vibrational model $\nu_{as}(C-N^+-C)$  at 970  $cm^{-1}$ still kept unchanged. This means that the positive center  $C-N^+-C$  was still not attacked by the anion ions due to the space screen effects of three methyl groups. The interesting thing is that the partial positive center $CO-O-C ^+$ vibrational model $\nu_{s}(CO-O-C ^+)$ at 1066  $cm^{-1}$  had moved to 1062  $cm^{-1}$ . The bending vibrational mode related to the two $(CH_2)_n$ tail of DOPC $\delta(CH_2)_n$ at 1465  $cm^{-1}$  had moved to 1418  $cm^{-1}$ and the intensity decreased a lot. This means that the partial positive centers $CO-O-C ^+$  were attacked by the anion ions and the two $(CH_2)_n$  tail might be attached with anion particles. Another characterization of the anion ions interaction with the vesicle surface is that the intensity of the partial positive center$CO-O-C ^+$  vibrational model $\nu_{s}(CO-O-C ^+)$ at 1062  $cm^{-1}$ now is larger than that of the negative center $PO_2^-$ vibrational models $ \nu_{s}(PO_2 ^-)$ at 1090  $cm^{-1}$ , which is the reverse case we saw before. Figure~\ref{fig:figure2}(c), however, is quite different from Figure~\ref{fig:figure2}(b), we can see that the vibrational bands that correspond to the negative center $PO_2^-$  vibrational models $\nu_{s}(PO_2 ^-)$ at 1244  $cm^{-1}$  had moved now only to 1239  $cm^{-1}$  and $ \nu_{s}(PO_2 ^-)$ at 1090  $cm^{-1}$  kept unchanged. But the difference between the intensity of asymmetric stretching vibration and the symmetric stretching vibration had increased a lot compare to the IR spectra of pure DOPC vesicles. This spectra characterization is very similar to that of $Ag^+$ ions interaction with the surface of the vesicles, which might mean a weak cation ions-DOPC surface interaction . We do observed the $(Ag_2Br)^+, (Ag_3Br_2)^+ and (Ag_4Br_3)^+ $ clusters formed in the same sample by MALTI mass spectrascopy~\cite{Zhang2000C, Zhang2019B}.   The positive center $C-N^+-C$ vibrational model $\nu_{as}(C-N^+-C)$ at 970 $cm^{-1}$ , partial positive center $CO-O-C ^+$  vibrational model $\nu_{s}(CO-O-C ^+)$ at 1066 cm-1 and the bending vibrational mode related to the two $(CH_2)_n$ tail of DOPC $\delta(CH_2)_n$  at 1465  $cm^{-1}$  are again kept unchanged within the resolution range. They are 971  $cm^{-1}$ , 1065 $cm^{-1}$ , and 1466  $cm^{-1}$ respectively. These mean that the positive center $C-N^+-C$, partial positive center $CO-O-C ^+$ may not been attacked by the anion ions and the two $(CH_2)_n$ tail may not been attached with particles. We might conclude that in the interaction between the $AgBr$ clusters and the surface of the DOPC vesicles, there might exist some positive charged clusters. No negative clusters were observed to interaction with the DOPC vesicles by IR.

Now it is interesting to answer the questions that why we can not observe the anion-DOPC vesicle interaction in the case of $CsF$ while in the case of DOPC vesicle solution of concentration 2mg/ml with 0.01M $Ag^+$ ions inside and 0.01M $Br^-$ outside vesicle prior to electroporation we do. As described before, the synthetic phospholipid DOPC has a zwittefionic head group. The head group is rotating and well ordered at surface of vesicles. There may exist a small size gap between each odered head group.  Since all the experiments had shown that no anions can reach the positive center $C-N^+-C$ of the head group during the space screen of three methyl group, the possible ions that may attack the surface of DOPC vesicles are the cations with small enough size which can go through the head channel to attack the negative center  $PO_2^-$ . In our case, $Ag^+$ ions satisfy all the conditions. That’s why we can observe the strong interactions between silver cations and the DOPC vesicles. $F^-$ ions may have suitable size, but they are anions. $K^+$ is cation, but its size may be too large to go through the head group gap channel. That is why both $F^-$ and $K^+$ have no obvious interactions with the DOPC vesicles.  The next interesting thing is to answer why in the case of DOPC vesicle solution of concentration 2 mg/ml with 0.01M $Ag^+$ ions inside and 0.01M $Br^-$ outside vesicle prior to electroporation we do observe the strong interaction between $Br^-$ ions and DOPC vesicles. From the procedure to prepare this solution we know that first we need prepare $Ag^+$  both inside and outside the vesicle. We know that silver ions can have strong interaction with the negative center $PO_2^-$   of head group of the DOPC. After this interaction, the ordered head groups may be changed due to the static electronic repulsion between occupied silver ions and the positive center $C-N^+-C$ of the head group. This may open a large gap between the two ordered head groups. This has been shown in Figure~\ref{fig:figure2}(a)middle insert picture. At this time , when the $Br^-$ ions solution add in, the $Br^-$ ions can go through the large gap to attack the partial positive center in the tail close to the head as it has been shown in Figure~\ref{fig:figure2}(a)middle insert picture. This is exactly what we see from the IR spectra as we described before.

\subsection{The Direct Laser Desorption (DLD) Mass Spectra  Studies}   

\begin{figure}[H]
\begin{center}
\epsfig{file=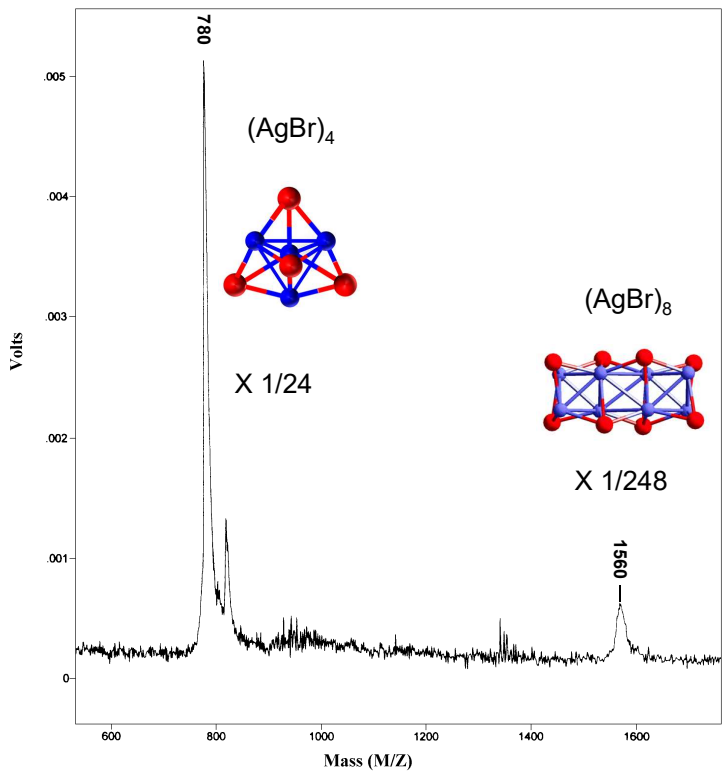,height=4.8in}
\caption{ DLDI mass spectra of silver bromide clusters showing the tetramer and Octamer.
}
\label{fig:figure3}
\end{center}
\end{figure}

\begin{table}[p]
\begin{center}

\begin{tabular}{p{1.6cm}p{1.8cm}p{2cm}p{1.0cm}p{4.0cm}p{2cm}} \hline 
MS exp value & Exp intensity &Calculated value & Merge peak & Isotopic formula & Cluster type\\ \hline 
- &  -   & 743.294   & 1 & \ce{^{107}Ag_4}\ce{^{79}Br_4}& $ (AgBr)_4$\\
- &  -   & 745.292 & 2 &\ce{^{107}Ag_4}\ce{^{79}Br_3}\ce{^{81}Br}&$ (AgBr)_4$\\
 &    & 745.293   &  & \ce{^{107}Ag_3} \ce{^{109}Ag}\ce{^{79}Br_4}& $ (AgBr)_4$\\

- &-    & 747.290   & 3 & \ce{^{107}Ag_4}\ce{^{79}Br_3}\ce{^{81}Br}& $ (AgBr)_4$\\
&    & 747.291   &  & \ce{^{107}Ag_3}\ce{^{109}Ag}\ce{^{79}Br_3}\ce{^{81}Br}&$ (AgBr)_4$\\
&    & 747.293   &  & \ce{^{107}Ag_2} \ce{^{109}Ag_2}\ce{^{79}Br_4}&$ (AgBr)_4$\\
- & -  & 749.288   & 4 & \ce{^{107}Ag_4}\ce{^{79}Br}\ce{^{81}Br_3}& $ (AgBr)_4$\\
&    & 749.289   &  & \ce{^{107}Ag_3}\ce{^{109}Ag}\ce{^{79}Br_2}\ce{^{81}Br_2}&$ (AgBr)_4$\\
&    & 749.291   &  & \ce{^{107}Ag_2}\ce{^{109}Ag_2}\ce{^{79}Br_3}\ce{^{81}Br}&$ (AgBr)_4$\\
&    & 749.293   &  & \ce{^{107}Ag}\ce{^{109}Ag_3}\ce{^{79}Br_4}& $ (AgBr)_4$\\
780 & 0.0051  & 751.286   & 5 & \ce{^{107}Ag_4}\ce{^{81}Br_4}& $ (AgBr)_4$\\
&    & 751.287   &  & \ce{^{107}Ag_3}\ce{^{109}Ag}\ce{^{79}Br}\ce{^{81}Br_3}&$ (AgBr)_4$\\
&    & 751.289   &  & \ce{^{107}Ag_2}\ce{^{109}Ag_2}\ce{^{79}Br_2}\ce{^{81}Br_2}&$ (AgBr)_4$\\
&    & 751.291   &  & \ce{^{107}Ag}\ce{^{109}Ag_3}\ce{^{79}Br_3}\ce{^{81}Br}& $ (AgBr)_4$\\
&    & 751.292   &  &\ce{^{109}Ag_4}\ce{^{79}Br_4}& $ (AgBr)_4$\\
- &  -   & 753.285   & 4 & \ce{^{107}Ag_3}\ce{^{109}Ag}\ce{^{81}Br_3}&$ (AgBr)_4$\\
 &     & 753.287   &  & \ce{^{107}Ag_2}\ce{^{109}Ag_2}\ce{^{79}Br}\ce{^{81}Br_3}&$ (AgBr)_4$\\
 &     & 753.289   &  & \ce{^{107}Ag}\ce{^{109}Ag_3}\ce{^{79}Br_2}\ce{^{81}Br_2}&$ (AgBr)_4$\\
 &     & 753.290   &  & \ce{^{109}Ag_4}\ce{^{79}Br_3}\ce{^{79}Br}&  $ (AgBr)_4$\\
- & -    & 755.285   & 3 & \ce{^{107}Ag_2} \ce{^{109}Ag_2}\ce{^{81}Br_4}&$ (AgBr)_4$\\
&    & 755.287   &  & \ce{^{107}Ag}\ce{^{109}Ag_3}\ce{^{79}Br}\ce{^{81}Br_3}& $ (AgBr)_4$\\
&    & 755.288   &  & \ce{^{107}Ag_4}\ce{^{79}Br_2}\ce{^{81}Br_2}& $ (AgBr)_4$\\
- &  -   & 757.285   & 2 & \ce{^{107}Ag}\ce{^{109}Ag_3}\ce{^{81}Br_4}&$ (AgBr)_4$\\
 &    & 757.286   &  & \ce{^{109}Ag_4}\ce{^{79}Br}\ce{^{81}Br_3}& $ (AgBr)_4$\\
- & -    & 759.284  &  1 & \ce{^{109}Ag_4}\ce{^{81}Br_4}&$ (AgBr)_4$\\
1560 & 0.0005  & 1502.571   & 9 & \ce{^{107}Ag_8}\ce{^{81}Br_8}& $ (AgBr)_8$\\
&    & 1502.573   &  & \ce{^{107}Ag_7}\ce{^{109}Ag}\ce{^{79}Br}\ce{^{81}Br_7}&$ (AgBr)_8$\\
&    & 1502.575   &  & \ce{^{107}Ag_6}\ce{^{109}Ag_2}\ce{^{79}Br_2}\ce{^{81}Br_6}&$ (AgBr)_8$\\
&    & 1502.576   &  & \ce{^{107}Ag_5}\ce{^{109}Ag_3}\ce{^{79}Br_3}\ce{^{81}Br_5}&$ (AgBr)_8$\\
&    & 1502.578   &  &\ce{^{107}Ag_4}\ce{^{109}Ag_4}\ce{^{79}Br_4}\ce{^{81}Br_4}& $ (AgBr)_8$\\
&    & 1502.580   &  & \ce{^{107}Ag_3}\ce{^{109}Ag_5}\ce{^{79}Br_5}\ce{^{81}Br_3}&$ (AgBr)_8$\\
&    & 1502.581   &  & \ce{^{107}Ag_2}\ce{^{109}Ag_6}\ce{^{79}Br_6}\ce{^{81}Br_2}&$ (AgBr)_8$\\
&    & 1502.583   &  & \ce{^{107}Ag}\ce{^{109}Ag_7}\ce{^{79}Br_7}\ce{^{81}Br}&$ (AgBr)_8$\\
&    & 1502.585   &  &\ce{^{109}Ag_8}\ce{^{79}Br_8}& $ (AgBr)_8$\\
 \hline
\end{tabular}
\caption{Comparison of the DLD mass peaks observed in the experiment with those calculated from the isotopic formula for all peaks of $ (AgBr)_4$ cluster and max merge peaks of $ (AgBr)_8$ cluster}
\label{tab:ISO}
\end{center}
\end{table}

The direct laser desorption (DLD) mass spectra were acquired using a home built linear time-of-flight (TOF) mass spectrometer in which samples are introduced through a hole in the center of the repeller electrode in the TOF source region~\cite{Kinsel}. Laser desorption is performed using an Oriel N2 laser(337nm) operating at a repetition rate of ca. 3Hz and adjusted to intensities ca. 10\% above threshold for analyte ion formation using a variable neutral density filter.

\begin{figure}[H]
\begin{center}
\epsfig{file=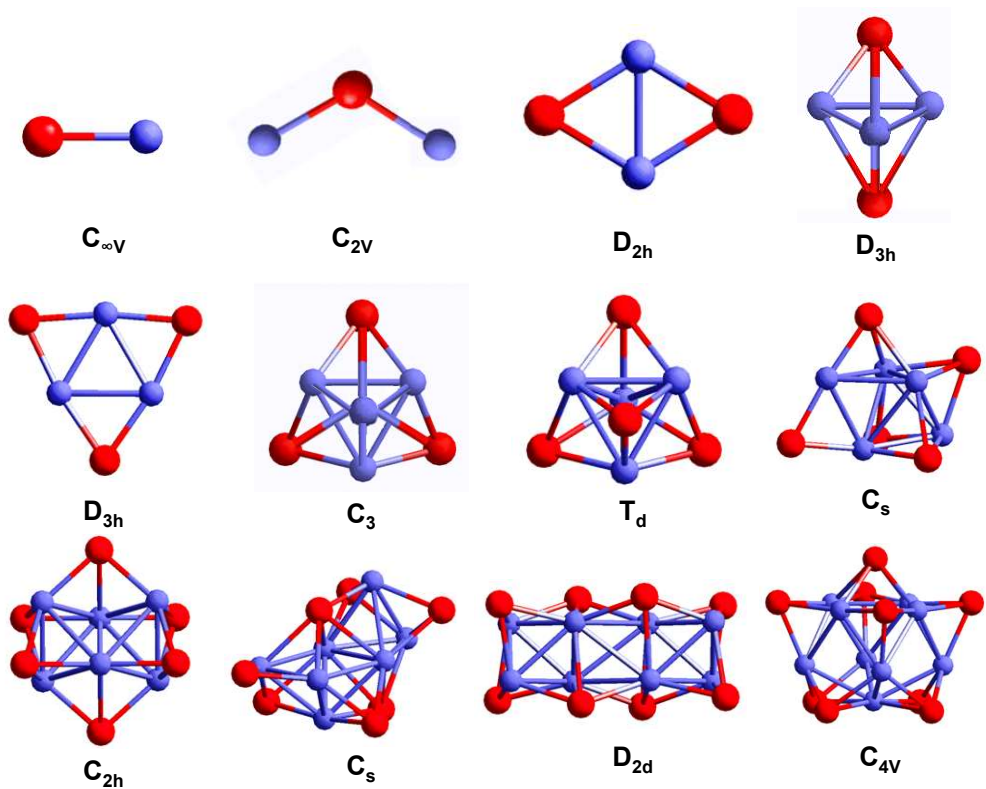,height=4.8in}
\caption{Silver bromide neutral and ion cluster structures with the lowest energies from the monomer to nanomer and their symmetries.}
\label{fig:figure4}
\end{center}
\end{figure}

In our previous work~\cite{Zhang2000C, Zhang2019B}, by using direct laser desorption mass spectra, we successfully observed  $(Ag_2Br)^+, (Ag_3Br_2)^+  and (Ag_4Br_3)^+ $ clusters in the same three samples that were used in the UV absorption experiments. For the first time, we experimentally confirm that the turn-around point cluster structure is $(Ag_3Br_2)^+$.  The DLD MS spectra of the three observed  $(Ag_2Br)^+, (Ag_3Br_2)^+ and (Ag_4Br_3)^+ $  clusters have an unsymmetrical cluster finger peaks and also the larger the clusters size, the lower DLD-TOF-MS spectra intensity. Also we can observe all the peaks predicted by the isotopic formula calculation for the small $(Ag_2Br)^+$ cluster but not for the large $(Ag_4Br_3)^+ $ cluster. All these can be explained by symmetry and probability principle in molecular cluster growth range we proposed in the previous work. 

Naturally occurring silver (Ag) is composed of two stable isotopes ${}^{107}Ag$ and ${}^{109}Ag$,  ${}^{107}Ag$ has 51.839\% natural abundance and 106.905097 u isotopic mass.  ${}^{109}Ag$  has 48.161\% natural abundance and 108.904752u isotopic mass. Naturally occurring bromine (Br) is also composed of two stable isotopes ${}^{79}Br$ and ${}^{81}Br$. ${}^{79}Br$ has 50.69\% natural abundance and 78.9183371 u isotopic mass.${}^{81}Br$ has 49.31\% natural abundance and 80.9162906u isotopic mass.  By considering these isotopic effects, all the peaks observed in the experiment can be assigned exactly within the error range of the experiment. For  $(Ag_nBr_{(n-1)}^+)$  ionic cluster, the number of predicted observe peak N is

\begin{equation}
 N=2n
\label{equ:a}
\end{equation}

the pattern of the number of predicted merge peak  is

\begin{equation}
 1:2:3:...:n-1:n:n:n-1:...:3:2:1
\label{equ:b}
\end{equation}

the number of overall peak $N_T$ is

\begin{equation}
 N_T=n(n+1)
\label{equ:c}
\end{equation}

For  $ Ag_nBr_n $  neutral cluster, the number of predicted observe peak N is

\begin{equation}
 N=2n+1
\label{equ:d}
\end{equation}

the pattern of the number of predicted merge peak  is

\begin{equation}
 1:2:3:...:n-1:n:n+1:n:n-1:...:3:2:1
\label{equ:e}
\end{equation}

the number of overall peak $N_T$ is

\begin{equation}
 N_T=(n+1)^2
\label{equ:f}
\end{equation}

In this work, we report the DLD MS spectra of sample that have UV absorption red-shift end point (273 nm) at large mass range. As shown in Figure~\ref{fig:figure3} and Table~\ref{tab:ISO}, we only observed one broad peak center at 780 for tetramer and one broad peak center at 1560 for octamer. It is reasonable to assign this center position to the calculated peak with largest peak merge number. For the tetramer, this is 751 with largest peak merge number 5 and for the octamer, this is 1502 with largest peak merge number 9. So there has a difference between the experiment data and the calculated data.  For the tetramer, the difference is 29 and for the octamer, the difference is 58. Where does this difference come from? We will discuss it in the next section.  The peak assignments have been shown in Table~\ref{tab:ISO}.     To our surprise, although our theoretical work predict there should have structure stable pentamer with $C_s$ symmetry, hexamer with $C_{2h}$ symmetry and heptamer with $C_s$ symmetry as shown in Figure~\ref{fig:figure4}, we didn't observe them. We only observed tetramer with $T_d$ symmetry and octamer with $D_{2d}$ symmetry with much lower intensity compare to $(Ag_2Br)^+$ cluster. Why? This will also be discussed in next section.

\subsection{Bi-layers Growth Mechanism of Silver Bromide Clusters Prepared Via Electroporation of Vesicles }
In molecular and small cluster size range, several growth mechanism of silver bromide clusters have been proposed. By using stopped-flow method, where the concentration of the bromide ions is much larger than the concentration of silver ions in the solution, both Tanaka~\cite{Tanaka1985A} and Ehrlich  ~\cite{Ehrlich} proposed the following polybromoargentate mechanism:\\
\begin{flushleft}
\ce{Ag+ + Br- ->AgBr}\\
\ce{AgBr + iBr- ->AgBr_{(i-1)}^{i-}}\\
\ce{nAgBr_{(i-1)}^{i-} -> Ag_nBr_{(n+j)}^{j-} + (ni-j)Br-}\\
\end{flushleft}
Here, n is the number of silver-bromide ion pairs needed for the formation of a primary particle (nucleus) of silver bromide, and was estimated to be 4. Symbol i represents the number of excess bromide ions present in the transient complex. Symbol j has been used to include the possibility that the particles are negatively charged due to adsorbed or bonded excess bromide ions (polynuclear complex anions). This mechanism is supported by the observation of the 230nm UV absorption band which was assigned to be the transient polybromoargentate complex anions $AgBr_i ^{(i-1)-}, (i=1,2,3,4)$  in both of their experiments. This can be a good example method to make the anion molecular clusters and we call it anion molecular cluster growth mechanism\\

By using the pulse radiolysis method ~\cite{Zhang1997A}, where the concentration of the bromide ions produced in the solution is almost the same as the concentration of silver ions in the solution, we had proposed the following growth mechanism:\\
\begin{flushleft}
\ce{Ag+ + Br- ->AgBr}\\
\ce{(AgBr)_{n-1} + Ag+ -> (Ag_nBr_{n-1})+}\\
\ce{(Ag_nBr_{n-1})+ + Br- -> (AgBr)_n}\\
\end{flushleft}

\begin{figure}[H]
\begin{center}
\epsfig{file=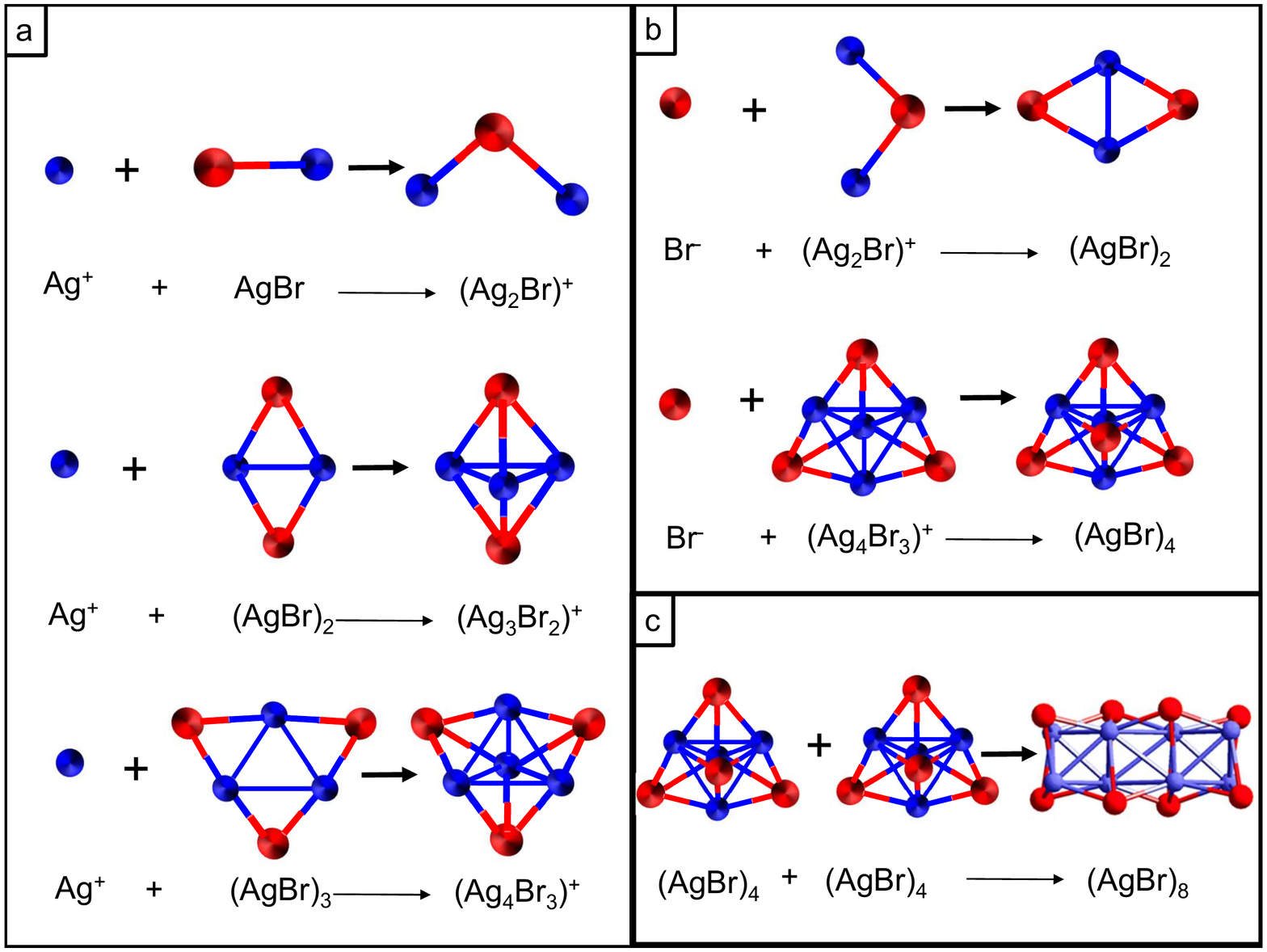,height=4.8in}
\caption{Diagram for the example reaction of (a)\ce{Ag+ + (AgBr)_n ->(Ag_{n+1}Br_n)+}.   (b) \ce{Br- + (Ag_{n+1}Br_n)+ ->(AgBr)_{n+1} } . (c) \ce{m(AgBr)_n -> (AgBr)_{mn}} .}
\label{fig:figure5}
\end{center}
\end{figure}

We reported the observation of AgBr monomer with the UV absorption band at 295nm and the AgBr dimer with the UV absorption band at 285nm ~\cite{Zhang1997A}. Based on our previous theoretical calculation of the neutral and ionic silver bromide cluster structure shown in Figure~\ref{fig:figure4}, this kind of growth mechanism can be described clearly as shown in Figure~\ref{fig:figure5}a and b. This can be a good example method to make the cation molecular clusters and we call it cation molecular cluster growth mechanism \\

Our experiments that prepared the AgBr clusters via the electroporation of vesicles let us  observed the entire blue-shift (274nm to 269nm) followed by red-shift (269nm to 273nm) of the absorption band that is associated with the growth of the silver bromide clusters. In this experiment, the following  growth mechanism was proposed\cite{Zhang2000A}:\\

\begin{flushleft}
\ce{Ag+ + Br- ->AgBr}\\
\ce{n(AgBr) -> (AgBr)_n}\\
\ce{m(AgBr)_n -> (AgBr)_{mn}}\\
\ce{m(AgBr)_n + l(AgBr)_p -> (AgBr)_{mn+lp}}\\
\end{flushleft} 

The third reaction equation is added based on the DLD-MS experimental observation that described in the previous section and an example of the third reaction has been shown in Figure~\ref{fig:figure5}c. The 4th reaction equation is a generalized reaction equation of the third one. Here $ m,n,l,p $ can be the integer number.  This can be a good example method to make the neutral molecular clusters and we call it neutral molecular cluster growth mechanism.  

\begin{figure}[H]
\begin{center}
\epsfig{file=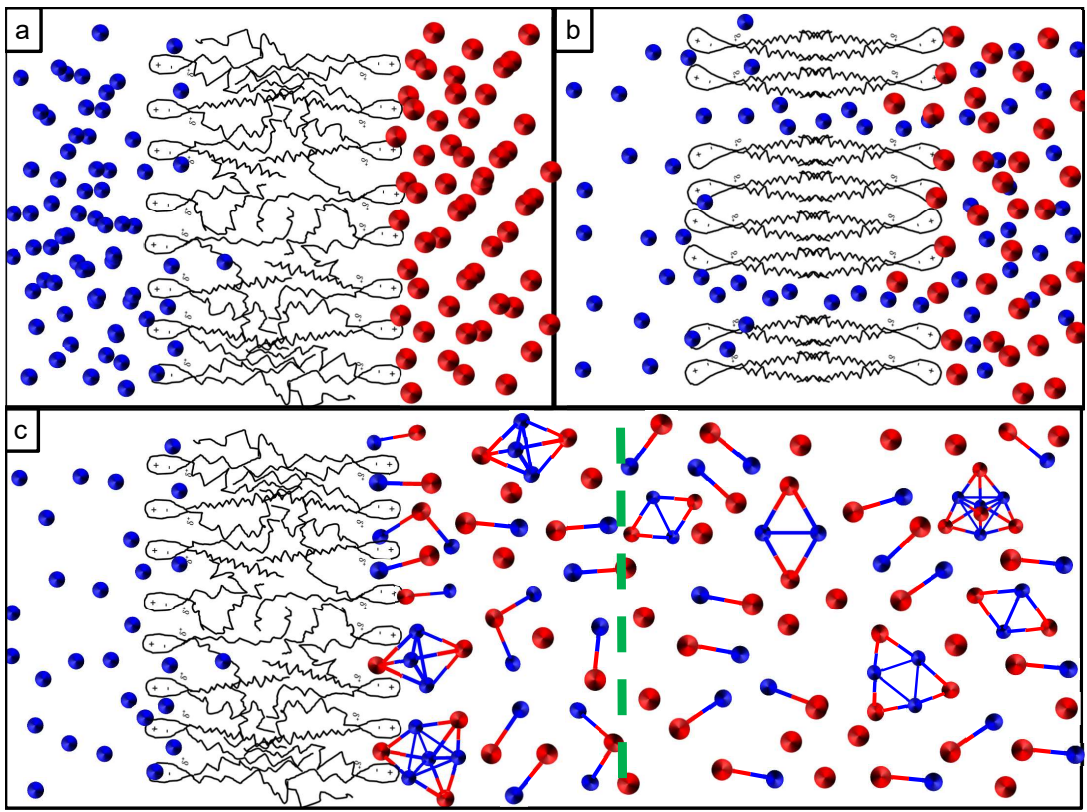,height=4.8in}
\caption{Enlarged local diagram of (a)DOPC vesicles containing entrapped $Ag^+$ ions, and $Br^-$ ions in the bulk prior to electroporation, (b)DOPC vesicles containing entrapped $Ag^+$ ions, and $Br^-$ ions in the bulk during the electroporation, (c)DOPC vesicles containing entrapped $Ag^+$ ions, and $Br^-$ ions in the bulk after electroporation and showing a distance dependent bilayer growth of silver bromide clusters outside the vesicles area.}
\label{fig:figure6}
\end{center}
\end{figure}

To understand those growth mechanisms that proposed in different experiment conditions and their UV absorption spectra behavior, we present the results of our theoretical studies on both neutral silver bromide clusters ($(AgBr)_n, n=1-9$)~\cite{Zhang2000B} and silver bromide ion clusters($ (Ag_nBr_{n-1})^+,   n=2-4$)~\cite{Zhang2000C, Zhang2012A}, the theoretically predicted stable silver bromide clusters and their symmetry have been described in Figure~\ref{fig:figure4}. In the previous section, our DLD-Mass spectra only observed  tetramer and octamer. This can be understood by the silver bromide cluster symmetry shown in Figure~\ref{fig:figure4}. Since in the tetramer to octamer mass range, tetramer and  octamer have the highest cluster symmetry. $ T_d $ and $D_{2d} $, based on the symmetry and probability principle in molecular cluster growth range: although pentamer with $C_s$ symmetry, hexamer with $C_{2h}$ symmetry and heptamer with $C_s$ symmetry  have high probability to form the clusters, but they are NOT stable, they can be easily decomposed to the more stable cluster tetramer with $ T_d $ symmetry plus other smaller clusters that have much higher probability  and stability. And two stable tetramer with $ T_d $ symmetry can then form a relative stable octamer with $D_{2d} $ symmetry as shown in Figure~\ref{fig:figure5}c. Another problem in the DLD-TOF-Mass spectra experiments is that there has a difference between the experiment data and the calculated data.  For the tetramer, the difference is 29 and for the octamer, the difference is 58. Where are the difference come from? This can be understood by the bi-layers growth mechanism of silver bromide clusters prepared via electroporation of vesicles as described in Figure~\ref{fig:figure6}. Figure~\ref{fig:figure6}a is a local enlarged diagram show that  entrapped $Ag^+$ ions in the DOPC membrane left side and outside $Br^-$ ions in the right side. Figure~\ref{fig:figure6}b describe during the electroporation, the DOPC membrane open the hole channel and allow the $Ag^+$ ions come out to react with $Br^-$ ions to form the silver bromide molecular clusters on the DOPC membrane surface according to the reaction equations described in the positive ion cluster growth mechanism. And this had be described in Figure~\ref{fig:figure5}a and b. This forms the first layer silver bromide clusters and we call it surface layer cluster formation. It is reasonable to assume some clusters formed in the surface layer can gain enough kinetic energy to go to the bulk solution and form larger clusters in the bulk and we call it bulk layer cluster formation. This had been described in Figure~\ref{fig:figure6}c. Based on our IR studies, we know that in our case, both silver and bromide ions can have strong interaction with DOPC. It is reasonable to assume during the cluster formation process,  the clusters bring some carbon residual from DOPC with it, then it will change the cluster mass spectra. In our case, we can assume the tetramer cluster bring a $CH_2CH_3$ residuals, which has a mass close to 29. Two these kinds of tetramers form the octamer will have 58 mass difference compare to pure octamer silver bromide clusters. This is exactly we observed in our DLD-Mass  spectra experiments.

Silver clusters formation inside AgBr clusters had been detected in our early experiments of surface enhanced Raman scattering of pyridine in AgBr Sol~\cite{Zhang1991A,Zhang1992A} . It was found that $Ag_4 $ cluster gives the largest contribution to the SERS intensity~\cite{Zhang1995A,Zhang1992A} and the intensity change can be used to study the nano AgBr cluster fractal aggregation structures~\cite{Zhang1994A,Zhang1992A}.  The DLD-MS studies of the larger silver bromide cluster formation agree well with the previous studies since both tetramer and octamer contain $Ag_4 $ clusters inside them.

\begin{figure}[H]
\begin{center}
\epsfig{file=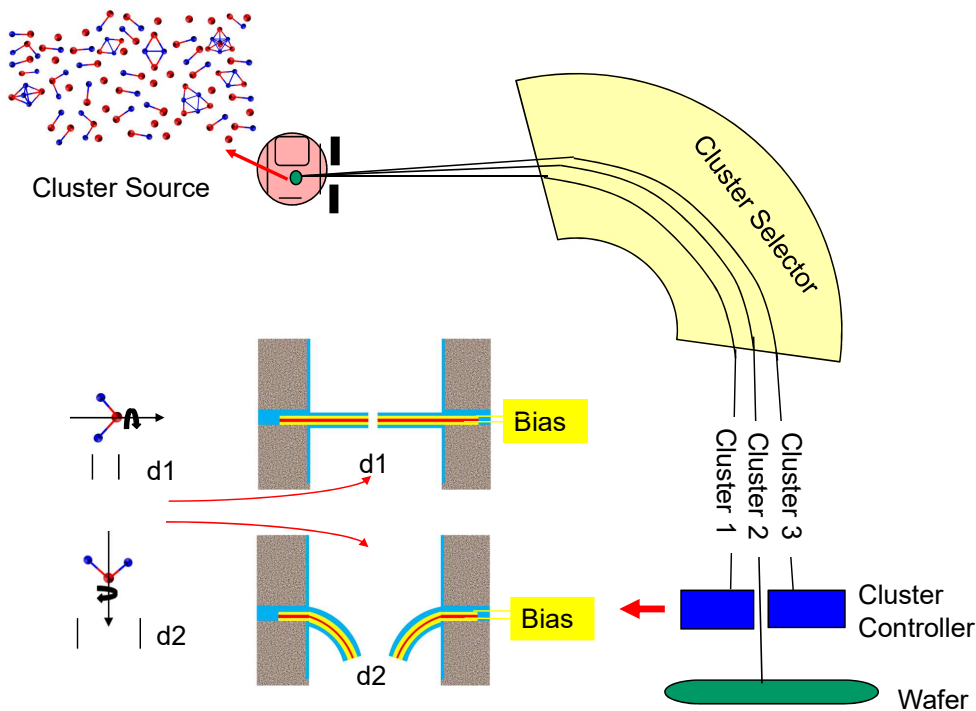,height=4.8in}
\caption{Diagram for selective molecular cluster deposition technology including cluster source, cluster selector and cluster controller  }
\label{fig:figure7}
\end{center}
\end{figure}
 
From the studies of Quasi-Elastic Light Scattering (QELS), FTIR, DLD-Mass spectra experiments, we know that cation and neutral molecular clusters can be prepared via the electroporation of vesicles. For the anion molecular clusters, it can be formed by the anion molecular formation mechanism we discussed above. The formation of the cation, anion and neutral molecular cluster source is a very important part for the selective molecular cluster deposition technology as shown in Figure~\ref{fig:figure7}. A beam of ions contains different ionic clusters which are then selected for deposition by passing the beam through a filter in which different apertures select clusters based on size and orientation~\cite{Zhang2018A,Zhang2015A}. From technology point of view, this paves the way for advanced device technology continue shrink and new concept device generation in the atomic and molecular cluster size range.

\section{Quantum Confinement Effects of Molecular Clusters and Their Applications }

\subsection{Quantum Confinement Effects of Molecular Clusters }

\begin{figure}[H]
\begin{center}
\epsfig{file=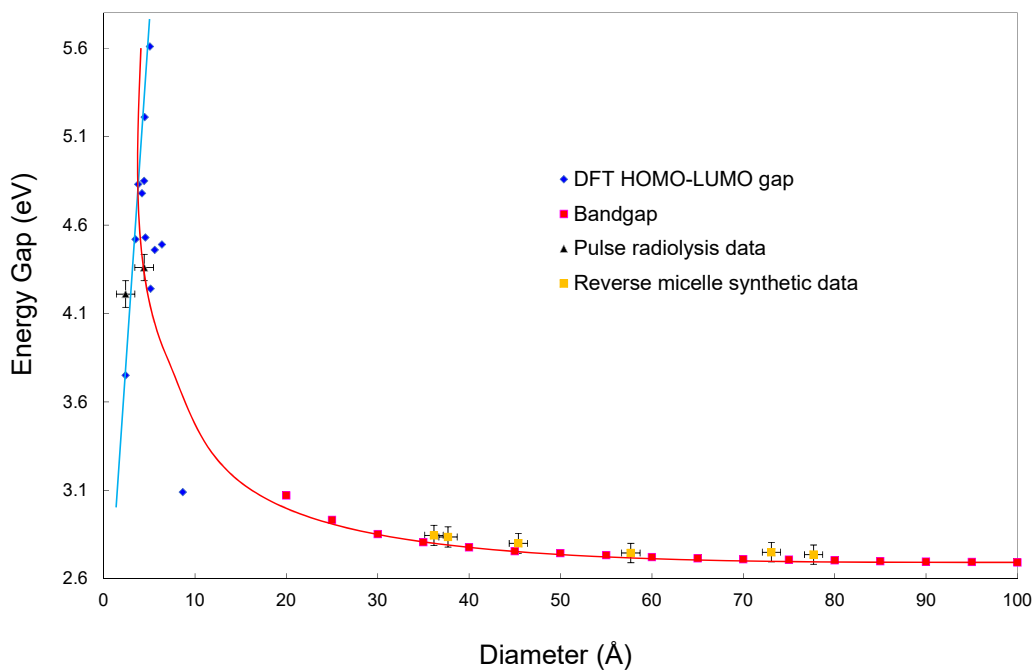,height=4.8in}
\caption{Experimental and theoretical data compare for the electronic absorption energy gap change vs cluster size  at molecular cluster range (blue line and dot for blue shift) and crystal cluster range (red line and dot for red shift). }
\label{fig:figure8}
\end{center}
\end{figure}

We show in our previous work that Quantum Confinement Effects of molecular clusters can be understood by a model that assume electrons are confined to a spherical potential well and the clusters are made of some basic units~\cite{Zhang2019A}. A formula is given for the lowest excited electronic state energy. 

\begin{equation}
 \ \Delta E= \left. \frac{\pi h^2 N_0}{6\mu V}\right. \bullet \zeta R
\label{equ:n}
\end{equation}

This expression contains an electron delocalization constant $ \zeta $. The same model without considering the electrons localization confinement, the lowest excited electronic state energy can be described by the following formula:

\begin{equation}
 \ \Delta E= E_{n+1}-E_n \approx  \left. \frac{n^2 h^2}{4\mu R^2}\right.  (\text{for } n \gg  1 )
\label{equ:k}
\end{equation}

\begin{figure}[H]
\begin{center}
\epsfig{file=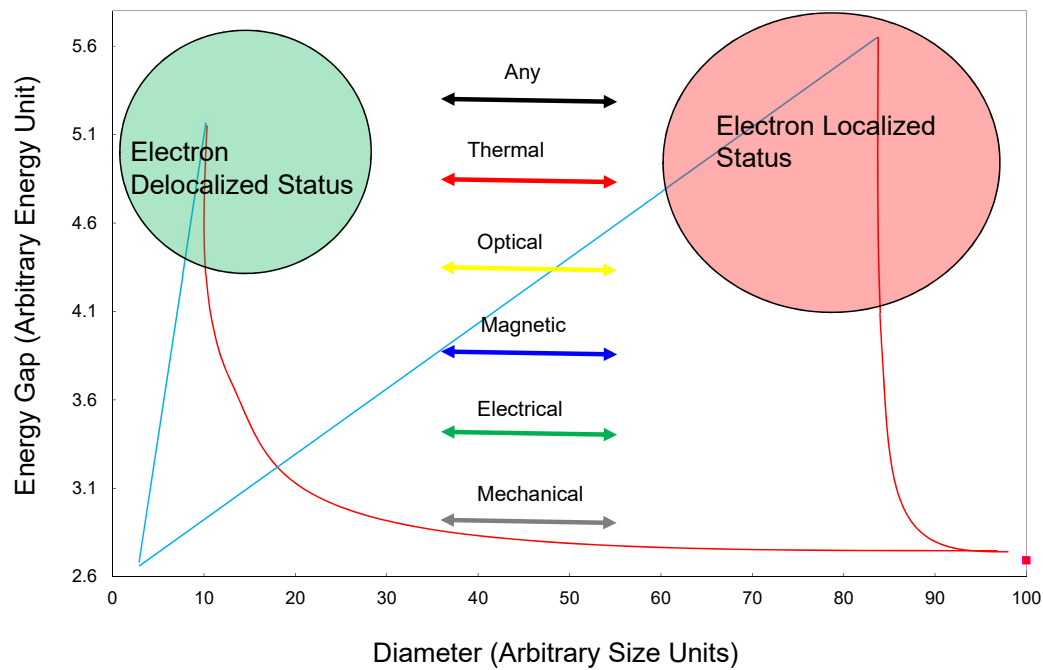,height=4.8in}
\caption{Diagram for molecular clusters switch between electron localized status and electron delocalized status by any kinds of interactions such as thermal, optical, magnetic, electrical and mechanical etc.   }
\label{fig:figure9}
\end{center}
\end{figure}

Here $ n $  is the quantum number of highest occupied orbital.  These two formulas describe  a $\lambda$ curve of $\Delta E$ vs sphere size $R$. As experimental method can enable to synthesize heavy monomer and dimer molecular clusters such as silver bromide clusters, this $\lambda$ curve had been proved to be a right curve to describe the molecular cluster quantum confinement behavior. Figure~\ref{fig:figure8} shows the experimental and theoretical data compare for the electronic absorption energy gap change vs cluster size  at molecular cluster range. The blue diamond dots are ab initio HF and density  functional theory (DFT) calculation results for AgBr neutral and ion clusters listed in Figure~\ref{fig:figure4}. Two blue triangle dots are pulse radiolysis experimental results for silver bromide monomer and dimers. Pink square dots are calculated from energy band theory and yellow square dots are reverse micelle synthetic experimental data. The blue line and red line are the fitting lines for these data and form exactly a $\lambda$ curve as we saw before. From previous studies we know that there can have two extreme status within the quantum confinement range $R$. Electron Delocalized Status (EDS), $ \zeta=1 $   when the electrons are completely delocalized and Electron Localized Status (ELS), $\zeta=0$  when the electrons are completely localized. We can define an EDS/ELS switch within quantum confinement range $R$ as following: Any interactions with the quantum confinement system within range $R$  that can cause the switches between EDS and ELS is called EDS/ELS switch or quantum confinement switch (QCS). The interactions can be thermal~\cite{Chang}, optical~\cite{Dudin}, magnetic~\cite{Natterer,Koepsell}, electric~\cite{Fuller,Valle} or mechanical~\cite{Eichler} etc as shown in Figure~\ref{fig:figure9}

The EDS/ELS switch or QCS pave the theoretical way for not only traditional technology node continue shrinking but also new concept transistor generation.

\subsection{Discussion of  Applications of Quantum Confinement Effects of Molecular Clusters }

\begin{figure}[H]
\begin{center}
\epsfig{file=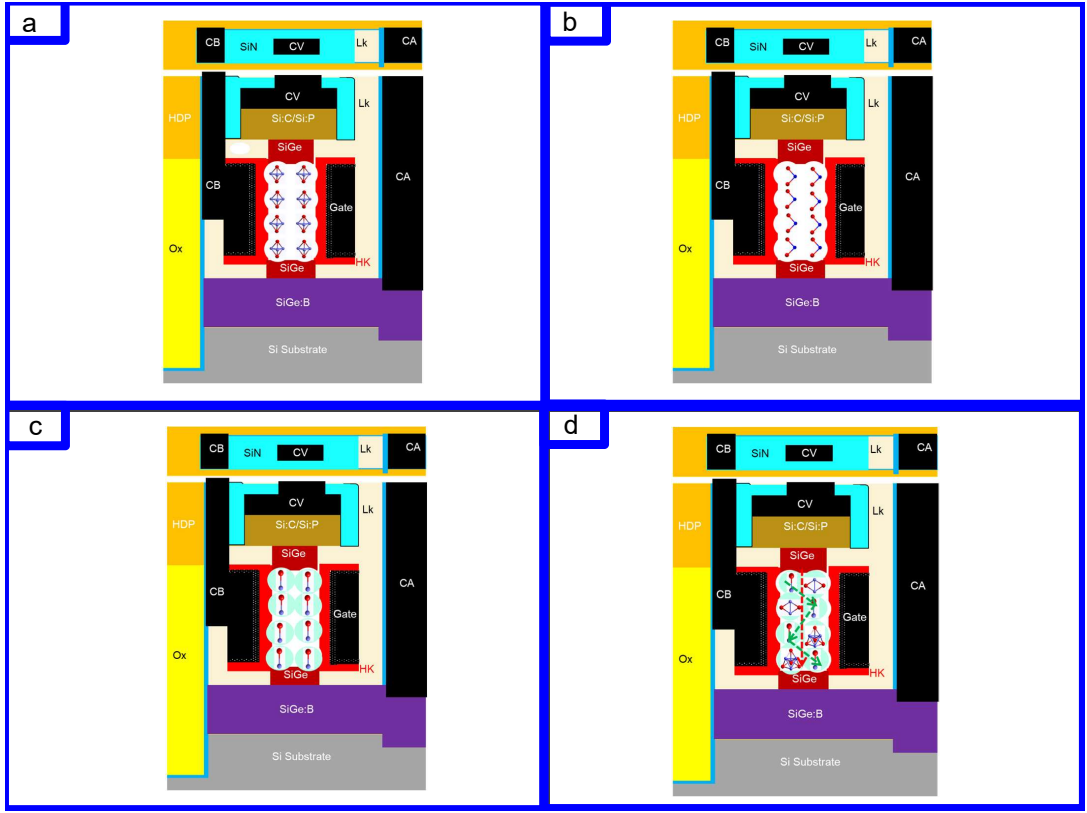,height=4.8in}
\caption{Device diagram for vertical TFET using (a) cation clusters for p channel transistor, (b)anion cluster for n channel transistor, (c) small neutral cluster for low Vt transistor and (d)mixing clusters channel for multi Vt and multi channel length transistor. }
\label{fig:figure10}
\end{center}
\end{figure}

Integrated circuits have various device architectures such as Vertical Field Effect Transistor (VFET)~\cite{Zhang2018B}, Vertical Turnal Field Effect Transistor(VTFET)~\cite{Zhang2018C,Zhang2016A}, Nanosheet Transistor~\cite{Zhang2017B, Zhang2018D,Zhang2017C}, Photonic Integrated Circuits~\cite{Zhang2015B, Zhang2019D}, Bio sensor~\cite{Zhang2017D}, 3D stack~\cite{Zhang2013A}, 3D package~\cite{Zhang2019E}, 3D cooling~\cite{Zhang2014A}, Radio Frequence Integrated Circuits~\cite{Zhang2019F}, Analog Integrated Circuits~\cite{Zhang2019F}, Logic Integrated Circuits~\cite{Zhang2019G}, Input/Output devices (I/O)~\cite{Zhang2017E}, Fully Deplete Silicon on Insulator device (FDSOI)~\cite{Zhang2018E}, Dynamic Random Access Memory (DRAM)~\cite{Zhang2015C}, Static Random Access Memory(SRAM)~\cite{Zhang2017F},  Phase Change Memory(PCM)~\cite{Zhang2015D}, Magnetic Random Access Memory (MRAM)~\cite{Zhang2018F}, Resistive Random Access Memory (RRAM)~\cite{Zhang2019H}, Vacuum transistor~\cite{Zhang2017G}, Thin Film Transistor~\cite{Zhang2018G},  Crack Stop~\cite{Zhang2017H}, Efuse~\cite{Zhang2016B}, Electical Static Diode(ESD)~\cite{Zhang2015E}, SiC device~\cite{Zhang2015F}, High Electron Mobility Transistor (HEMT)~\cite{Zhang2015G} and micro-electromechanical system (MEMS) devices~\cite{Zhang2016C} etc.

\begin{figure}[H]
\begin{center}
\epsfig{file=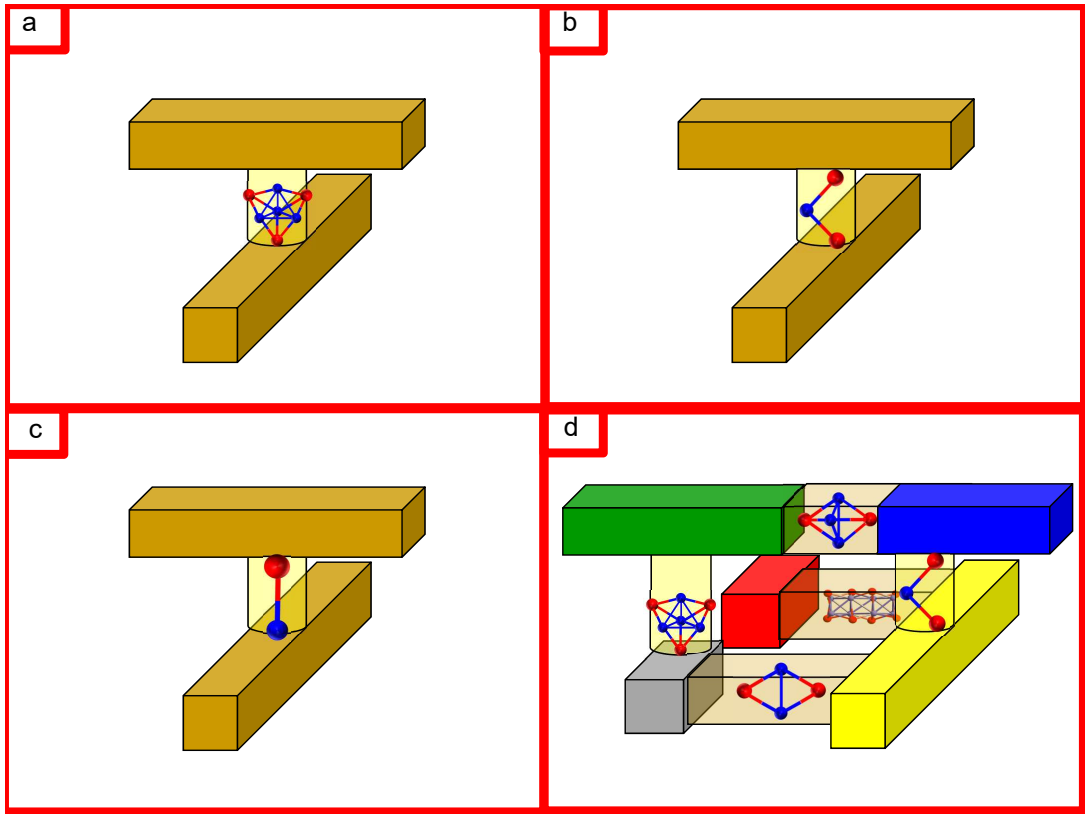,height=4.8in}
\caption{Device diagram for new concept transistor doing local and delocalization switch using (a) cation cluster, (b) anion cluster, (c) different size neutral cluster and (d) mixing clusters with multi different type connection.  }
\label{fig:figure11}
\end{center}
\end{figure}

As technology node continues to shrink, vertical transistor becomes one of the most attractive post-Fin Field Effect Transistor (FinFET) device architectures. However to implement the multi throsheld voltage on the vertical devices become extrme charlenge since the requriement of deposition and strip of multi different work function material layers within limited small space~\cite{Zhang2017A}.    With EDS/ELS switch or QCS, this can be solved easily, for the p channel device, just confine the cation molecular clusters within the channel, using gate to control the EDS/ELS switch, when it is in the EDS, the top and bottom electrode will get connected, when in the ELS, the top and bottom electrode connection will be broken. Same mechanism can be used to build n channel devices with anion clusters confined in the channel, different size neutral or ion cluster confined in the channel for Vt adjustment and mixing type clusters confined in  the channel for multi Vt and multi channel length transistor as shown in Figure~\ref{fig:figure10}.

EDS/ELS switch or QCS can also be used to develop new concept of transistors in the molecular cluster size range, this has been shown in Figure~\ref{fig:figure11}. Figure~\ref{fig:figure11}a shows a EDS/ELS switch or QCS new concept transistor realized by only one cation cluster confined in a via structure. This can be manufactured by the selective molecular cluster process using a hole size controllable device as shown in Figure~\ref{fig:figure7} and described in previous sections~\cite{Zhang2018A,Zhang2015A}. Figure~\ref{fig:figure11}b and c show a EDS/ELS switch transistor with confined anion and neutral molecular clusters. Figure~\ref{fig:figure11}d shows a new concept EDS/ELS switch or QCS transistor switched by different type interactions. The red contacts is for thermal, the yellow is for optical, the blue is for magnetic, the green is for electric and the gray is for mechanical etc interaction switch as described in Figure~\ref{fig:figure9} and previous section.

\section{Conclusion}
The studies of QELS, FTIR, DLD-Mass spectra experiments show that cation and neutral molecular clusters can be prepared via the electroporation of vesicles.  The symmetry and probability principles in molecular cluster growth range was used to explain the lager silver bromide molecular clusters tetramer and octamer formation in a bilayer formation mechanism. Both DLD-MS studies of the larger silver bromide cluster formation and previous studies reveal that the $Ag_4 $ clusters play an important rule during the larger silver bromide clusters growth. The anion molecular clusters formation mechanism also discussed in this paper. The formation of the cation, anion and neutral molecular cluster source is a very important part for the selective molecular cluster deposition technology.  The $\lambda$ curve had been proved to be a right curve to describe the molecular cluster quantum confinement behavior. We also defined EDS,  ELS, and EDS/ELS switch or QCS in the quantum confine system. These studies pave the ways for advanced device technology continue shrink and new concept device generation in the atomic and molecular cluster size range.


\begin{thebibliography}{99}

\bibitem{Rossetti}

Rossetti R., Nakahara S. \&  Brus L. E., Quantum size effects in the redox potentials. resonance Raman spectra, and electronic spectra of .CdS crystallites in aqueous  solution. J. Chem. Phys., 79, 1086-1088(1983).

\bibitem{Onushchenko}
Ekimov A. I. \& Onushchenko A. A.,  Quantum size effect in three-dimensional microscopic semiconductor crystals.  JETP letter, 34, 345-348(1981).
\bibitem{Scholl}
Scholl J. A., Koh A. L. \& Dionne J. A., Quantum plasmon resonances of individual metallic nanoparticles. Nature. 483, 421–427 (2012).
\bibitem{Guo}
Guo Y.  et al., Superconductivity Modulated by Quantum Size Effects. Science 306, 1915-1917 (2004).
 
\bibitem{Cullis} 
Cullis A. G.  \& Canham L. T.,  Visible light emission due to quantum size effects in highly porous crystalline silicon. Nature, 353, 335-338(1991).

\bibitem{Konstantatos}
Konstantatos G.  et al., Ultrasensitive solution-cast quantum dot photodetectors.  Nature , 442, 180-183 (2006).
\bibitem{Behrens}
Behrens M. et al., The Active Site of Methanol Synthesis over Cu/ZnO/Al2O3 Industrial Catalysts. Science , 336, 893-897 (2012)
\bibitem{Zhang2019A}
Zhang John H,  Quantum Confinement Effects for Semiconductor Clusters in the Molecular Regime.  arXiv:1904.03666 [physics.chem-ph],2019

\bibitem{Zhang2019B}
Zhang John H,  Studies of Silver Bromide Clusters Isotopic Properties and Their Applications.  arXiv:1905.11315 [physics.chem-ph],2019

\bibitem{Alivisatos} 
Alivisatos A. P., Semiconductor clusters, nanocrystals, and quantum dots. Science, 271, 933-937(1996).
 
\bibitem{Alivisatos1}
Alivisatos A. P., Perspectives on the physical chemistry of semiconductor nanoctystals. J. Phys. Chem., 100, 13226-13239(1996).
 
\bibitem{Chen1997A}
Chen C. C., Herhoid A. B., Johnson C. S. \& Alivisatos A. P., Size dependence of structural metastability in semiconductor nanocrystals. Science, 276, 398-401(1997).

\bibitem{James}
James T. H.,  The Theory of the Photographic Process (Macmillan, 1977).

\bibitem{Zhang2017A}
Zhang John H., US9748356B2 2017-08-29 Threshold adjustment for quantum dot array devices with metal source and drain

\bibitem{Zhang2019C}
Zhang John H.,  US10199505B2 2019-02-05 Transistors incorporating metal quantum dots into doped source and drain regions

\bibitem{Tadaaki}
Tadaaki T.,  Photographic Sensitivity (Oxford University Press,1995).
 
\bibitem{Ehrlich}
Ehrlich S. H., The kinetic processes of formation and electron-trapping efficiencies of quantum-sized silver bromide clusters.  J.Imaging Sci.\&Tech., 38, 201-216(1994).
 
\bibitem{Chen1994A}
Chen W. et al.,  Size dependence of radiative rates in the indirect band gap material AgBr. J. Am. Chem. Soc., 116, 1585-1586 (1994).
 
\bibitem{Masumoto} 
Masumoto Y., Kawamura, T., Ohzeki T.\& Urabe S., Lifetime of indirect   excitons in  AgBr quantum dots. Phys. Rev. B., 46,1827-1830(1992). 
 
\bibitem{Johannson}
Johannson K. P., McLendon G. L. \& Marchetti A. P., The effect of size restriction on silver bromide. A dramatic enhancement of free exciton luminescence. Chem. Phys. Lett., 179, 321-324(1991).
 
\bibitem{Marchetti}
Marchetti A. P., Johannson K. P. \& McLendon G. L., AgBr photophysics from optical studies of quantum confined crystals. Phys. Rev. B, 47, 4268-4275(1993).
 
\bibitem{Kanzaki}
Kanzaki H. \& Tadakuma Y., Indirect exciton confinement and impurity isolation in  ultrafine particles of silver bromide. Solid State Commun., 80, 33-36(1991).
 
\bibitem{Freedhoff}
Freedhoff M. I., Marchetti A. P. \& McLendon G. L., Optical properties of nanocrystalline silver halides. J. Luminescence, 70, 400-413(1996).
 
\bibitem{von}
von der Osten W. \& Stolz H., Localized exciton states in silver halides. J. Phys. Chem. Solids, 51, 765-791(1990).
 
\bibitem{Scholle2}

Scholle U., Stolz H. \& von der Osten  W., Resonant Raman scattering and luminescence from size-quantized indirect exciton states in AgBr microcrystals.  Solid State Commun., 86, 657-661(1993).
 
\bibitem{Vogelsang}

Vogelsang H., Stolz H. \& von der Osten, W., Exciton-phonon coupling in indirect gap AgBr nanocrystals. J. Luminescence, 70, 414-420(1996).
 
\bibitem{Tanaka}

Tanaka T., Matsubara T., Saeki H.\& Heda H., Optical absorption studies of the growth of miccrocrystals in nascent suspensions. I. Silver chloride hydrosols.  Photogr. Sci. Eng., 20, 213-219(1976).
 
\bibitem{He}

He T., Wang P., Zhang X. \& Liu F., QSE of nanomer size semiconductor AgBr,  Chin. J. Chem. Phys., 8, 23-27(1995).
 
\bibitem{Zhang1997A}

Zhang H. \& Mostafavi M., UV-Absorption Observation of the Silver Bromide Growth from a Single Molecule to the Crystal in Solution. J. Phys. Chem. B, 101, 8443-8448 (1997).
 
\bibitem{Brus}

Brus L. E., Electron-electron and electron-hole interactions in small semiconductor crystallites. J. Chem. Phys., 80, 4403-4409(1984).
 
\bibitem{Chen1994B}

Chen W. et al.,  Luminnescence properties of indirect bandgap    semiconductors: nanocrystals of silver bromide. Mot. Clyst. Lig. Cryst., 252, 79-86(1994).
 
\bibitem{Zhang2000A}

Correa N. M., Zhang H. \& Schelly Z. A., Preparation of AgBr quantum dots via electroporation of vesicles. J. Am. Chem. Soc., 122, 6432-6434(2000).
 
\bibitem{Tanaka1}

Tanaka T., Saijo H. \&  Matsubara T., Optical absorption studies of the growth of microcrystals in nascent suspensions. III Absorption spectra of nascent iodide hydrosols. J. Photogr. Sci., 27, 60-65(1979)
 
\bibitem{Zhang2000B}

Zhang H., Schelly Z. A., \& Marynick D. S. Theoretical Study of the Molecular and Electronic Structures of Neutral Silver Bromide Clusters (AgBr)n, n = 1-9. J. Phys. Chem. A, 104, 6287-6294(2000).

\bibitem{Zhang2000C}
Zhang Hongguang,  Theoretical and experimental studies of silver bromide clusters. Thesis (PhD). The University of Texas at Arlington, Source DAI-B 61/05, p. 2556, Nov 2000, 115 pages. http://adsabs.harvard.edu/abs/2000PhDT........18Z

\bibitem{Zhang2012A}
 Zhang John H, Zoltan A. Schelly, Dennis S. Marynick \& Gary R. Kinsel, Quantum size effect studies of isotopic and electronic properties of silver bromide ionic clusters. Nature, September, 2012, unpublished.        

\bibitem{Zhang2018A}
Zhang John H.,  US10002938B2 2018-06-19 Atomic layer deposition of selected molecular clusters.

\bibitem{Zhang2015A}
Zhang John H.,  US9214622B2 2015-12-15 Size-controllable opening and method of making same.

\bibitem{Minami}
Hideyuki Minami \& Tohru Inoue, Aggregation of Dipalmitoylphosphatidylcholine Vesicles Induced by Some Metal Ions with High Activity for Hydrolysis. Langmuir  15, 6643-6651(1999). 

\bibitem{Korchowiec}
Korchowiec B., Puggelli, M., Gabrielli G.,  Nocentini M. \&  Focardi C., Thermodynamic and spectroscopic properties of mixtures of $\beta$ Mactoglobulin and dioleylphosphatidylcholine. Colloid Polym Sci, 275, 860-868 (1997). 

\bibitem{Chung}
Chung J. B.,  Hanneman, R. E. \& Franses E. I., Surface analysis of lipid layers at the A/W interface. Langmuir 6, 1647–1655(1990).

\bibitem{Meuse}
Meuse C.W., Niaura G., Lewis M.L. \& Plant A.L., Assessing the Molecular Structure of Alkanethiol Monolayers in Hybrid Bilayer Membranes with Vibrational Spectroscopies. Langmuir,  14, 1604-1611(1998).

\bibitem{Akutsu1981}
Akutsu H., Direct determination by Raman scattering of the conformation of the choline group in phospholipid bilayers. Biochemistry, 20, 7359-7366(1981).

\bibitem{Akutsu1986}
Akutsu H., Suezaki Y., Yoshikawa W. \&  Kyogoku Y.,Influence of metal ions and a local anesthetic on the conformation of the choline group of phosphatidylcholine bilayers studied by Raman spectroscopy.  Biochim. Biophys. Acta, 854, 213-218(1986). 

\bibitem{Kinsel}
Gimon-Kinsel M., Preston-Schaffter L. M., Kinsel G. R. \& Russell D. H., Effects of Matrix Structure/Acidity on Ion Formation in Matrix-Assisted Laser Desorption Ionization Mass Spectrometry. J. Am. Chem. Soc., 119, 2534–2540(1997).


\bibitem{Tanaka1985A}
Tanaka T. \& Iwasaki M. The multistage process of formation of ultrafine silver bromide particles as revealed by multichannel spectrophotometry. J. Imaging Sci., 29, 86-92 (1985).

\bibitem{Zhang1991A}
Zhang H, Xin H., He T. \& Liu F. Time dependent surface enhanced Raman spectroscopy of pyridine in AgBr sol. Spectrochimica  Acta A , 47, 927-932(1991).

\bibitem{Zhang1992A}
Zhang  Hongguang, Studies on Topological and Fractal Properties of High Tc Superconductors and  Colloids. Thesis (PhD). University of Science and Technology of China, May 1992, 118 pages, 108.ndlc.2.1100009031010001/T3F24.012002631352,O511.

\bibitem{Zhang1995A}
Zhang H, Liu F., He T. \& Xin H., Time dependent UV-absorption spectra and surface enhanced Raman spectroscopy of pyridine in AgCl sol. Spectrochimica  Acta A, 51, 1903-1908(1995).

\bibitem{Zhang1994A}
Zhang H., Liu F., He T. \& Xin H,  A study of relationship between universality in fractal colloids aggregation and time dependent SERS. Science in China  B, 37, 395-401(1994).


\bibitem{Chang}
Cheng Chang et al., 3D charge and 2D phonon transports leading to high out-of-plane ZT in n-type SnSe crystals.  Science, 360, 778–783 (2018)

\bibitem{Dudin}
Dudin Y. O. \& Kuzmich A., Strongly Interacting Rydberg Excitations of a Cold Atomic Gas. Science, 336, 887-889(2012). 

\bibitem{Natterer}
Natterer Fabian D. et al.,Reading and writing single-atom magnets. Nature, 543, 226-228(2017).

\bibitem{Koepsell}
Koepsell Joannis  et al., Imaging magnetic polarons in the doped Fermi–Hubbard model. Nature, 572, 358-362(2019).

\bibitem{Fuller}
Fuller Elliot J. et al., Parallel programming of an ionic floating-gate memory array for scalable neuromorphic computing. Science, 364, 570–574 (2019).

\bibitem{Valle}
Valle Javier del et al.,Subthreshold firing in Mott nanodevices. Nature, 569, 388-392(2019).

\bibitem{Eichler}
Eichler Alexander, Little is lost-Nanostrings can exploit strain engineering for unprecedented mechanical performance. Science, 360, 706–707(2018).

\bibitem{Zhang2018B}
Zhang John H.,  US9882025B1 2018-01-30 Methods of simultaneously forming bottom and top spacers on a vertical transistor device.

\bibitem{Zhang2018C}
Zhang John H.,  US9997463B2 2018-06-12 Modular interconnects for gate-all-around transistors.


\bibitem{Zhang2016A}
Zhang John H.,  US9385195B1 2016-07-05 Vertical gate-all-around TFET.


\bibitem{Zhang2017B}
Zhang John H., Cheng Kangguo, Clevenger Lawrence A \& Pranatharthiharan Balasubramanian S, US20170330934A1 2017-11-16 Devices and methods of forming self-aligned uniform nano sheet spacers.


\bibitem{Zhang2018D}
Cheng Kangguo, Clevenger Lawrence A, Pranatharthiharan Balasubramanian S \& Zhang John H., US20180114834A1 2018-04-26 Nanosheet transistors with sharp junctions.


\bibitem{Zhang2017C}
Cheng Kangguo, Clevenger Lawrence A, Pranatharthiharan Balasubramanian S \& Zhang John H., US9660028B1 2017-05-23, Stacked transistors with different channel widths.

\bibitem{Zhang2015B}
Zhang John H., US9116319B2 2015-08-25 Photonic integrated circuit having a plurality of lenses.

\bibitem{Zhang2019D}
Zhang John H., US10247881B2 2019-04-02 Hybrid photonic and electronic integrated circuits.



\bibitem{Zhang2017D}
Zhang John H.,  US9730596B2 2017-08-15 Low power biological sensing system.

\bibitem{Zhang2013A}
Zhang John H.,  US8564137B2 2013-10-22 System for relieving stress and improving heat management in a 3D chip stack having an array of inter-stack connections.



\bibitem{Zhang2019E}
Lawrence A. Clevenger, Carl Radens,  Yiheng Xu \& Zhang John H.,  US10304815B2 2019-05-28 Self-aligned three dimensional chip stack and method for making the same.


\bibitem{Zhang2014A}
Zhang, John H, Clevenger, Lawrence A, Radens, Carl, Yiheng, XU and  Wornyo, Edem  US8829670B1 2014-09-09 Through silicon via structure for internal chip cooling.


\bibitem{Zhang2019F}
Zhang,John H. US10249568B2 2019-04-02 Method for making semiconductor device with stacked analog components in back end of line (BEOL) regions.

\bibitem{Zhang2019G}
Liu Qing \& Zhang John H., US10319647B2 2019-06-11 Semiconductor structure with overlapping fins having different directions, and methods of fabricating the same.


\bibitem{Zhang2017E}
Wu Xusheng,Zhang John H. \& Huang Haigou,  US9837553B1 2017-12-05 Vertical field effect transistor.


\bibitem{Zhang2018E}
Zhang John H., US9947772B2 2018-04-17 SOI FinFET transistor with strained channel.

\bibitem{Zhang2015C}
Zhang John H.,  US20150318285A1 2015-11-05  Dram interconnect structure having ferroelectric capacitors.


\bibitem{Zhang2017F}
Zhang John H., US9825055B2 2017-11-21 FinFETs suitable for use in a high density SRAM cell.

\bibitem{Zhang2015D}
Zhang,John H., US9136473B2 2015-09-15 Semiconductor device with PCM memory cells and nanotubes and related methods.


\bibitem{Zhang2018F}
Zhang,John H.,  US10128327B2  2018-11-13 DRAM interconnect structure having ferroelectric capacitors exhibiting negative capacitance.


\bibitem{Zhang2019H}
Liu, Qing \& Zhang,John H.,  US10211257B2  2019-02-19 High density resistive random access memory (RRAM)


\bibitem{Zhang2017G}
Zhang John H.,  US9853163B2 2017-12-26 Gate all around vacuum channel transistor.


\bibitem{Zhang2018G}
Clevenger Lawrence A., Radens Carl, Xu Yiheng  \& Zhang John H., US10026849B2 2018-07-17 Structure and process for overturned thin film device with self-aligned gate and SD contacts.

\bibitem{Zhang2017H}
Zhang John H, Clevenger Lawrence A., Radens Carl, Xu Yiheng, Kim Youp  \& Kleemeier Walter ,  US9646939B2 2017-05-09 Multilayer structure in an integrated circuit for damage prevention and detection and methods of creating the same


\bibitem{Zhang2016B}
Zhang John H, Clevenger Lawrence A., Radens Carl, Xu Yiheng \& Wornyo Edem,  US9240375B2 2016-01-19 Modular fuses and antifuses for integrated circuits.


\bibitem{Zhang2015E}
Zhang John H.,Clevenger Lawrence A, Radens Carl \&  Xu Yiheng,  US8970004B2 2015-03-03 Electrostatic discharge devices for integrated circuits.

\bibitem{Zhang2015F}
Zhang John H. \& Morin Pierre, US9224845B1 2015-12-29 Silicon carbide static induction transistor and process for making a silicon carbide static induction transistor.

\bibitem{Zhang2015G}
Zhang John H.,  Kleemeier Walter \& Goldberg Cindy,  US8987780B2 2015-03-24 Graphene capped HEMT device.


\bibitem{Zhang2016C}
Liu Qing \& Zhang John H., US9466452B1 2016-10-11 Integrated cantilever switch.



\end{thebibliography}
\end{document}